\documentclass{ifacconf}

\usepackage{graphicx}      
\usepackage{natbib}        

\usepackage{mathtools}
\usepackage{amssymb}
\usepackage{pifont}
\DeclareMathOperator*{\argmin}{arg\,min}
\usepackage{multirow}
\usepackage{graphicx}
\usepackage{epsfig}
\usepackage{color}
\usepackage{dsfont}
\usepackage{soul}

\usepackage[linesnumbered,ruled,algo2e]{algorithm2e}
\DeclareMathAlphabet\mathbfcal{OMS}{cmsy}{b}{n}

\def\trace{\mathrm{tr}}

\def\Sigmahat{\widehat{\Sigma}}

\def\Sigbold{\boldsymbol{\Sigma}}

\def\muhat{\hat{\mu}}

\def\thetabold{\boldsymbol{\theta}}

\def\mubold{\boldsymbol{\mu}}

\def\ones{ \mathds{1} }
\def\wbold{\boldsymbol{w}}

\def\DKL{D_\textit{KL}}

\def\DL2{D_{\textit{L}2}}

\def\DJR2{D_\textit{JRD}^2}

\def\DW2{D_{\textit{W}2}}
\def\DH2{D_{\textit{H}2}}

\def\Real{\mathbb{R}}

\def\vec{\textit{vec}}

\def\nuhat{\hat{\nu}}

\def\Bcal{\mathcal{B}}

\def\nubold{\boldsymbol{\nu}}

\def\Lcaltilde{\tilde{\mathcal{L}}}

\def\qbold{\boldsymbol{q}}

\def\Bbreve{\breve{B}}

\def\d{\mathrm{d}}

\begin{document}
\begin{frontmatter}

\title{Adaptive mixture approximation for target tracking in clutter} 


\author[First]{Alessandro D'Ortenzio} 
\author[First]{Costanzo Manes} 
\author[Second]{Umut Orguner}

\address[First]{Department of Information Engineering, Computer Science and Mathematics, University of L'Aquila, Center of Excellence EX-Emerge,\\ (e-mail: alessandro.dortenzio@univaq.it, costanzo.manes@univaq.it)}
\address[Second]{Department of Electrical and Electronics Engineering, Middle East Technical University, Turkey, (e-mail: umut@metu.edu.tr)}

\begin{abstract}
Target tracking represents a state estimation problem recurrent in many practical scenarios like air traffic control, autonomous vehicles, marine radar surveillance and so on. In a Bayesian perspective, when phenomena like clutter are present, the vast majority of the existing tracking algorithms have to deal with association hypotheses which can grow in the number over time; in that case, the posterior state distribution can become computationally intractable and approximations have to be introduced. In this work, the impact of the number of hypotheses and corresponding reductions is investigated both in terms of employed computational resources and tracking performances. For this purpose, a recently developed adaptive mixture model reduction algorithm is considered in order to assess its performances when applied to the problem of single object tracking in the presence of clutter and to provide additional insights on the addressed problem.
\end{abstract}

\end{frontmatter}

\section{Introduction}
Reconstructing the state of a system, through a discrete-time sequence of noisy observations, is a problem which dates far back in time. Although many solutions have been proposed, one of the most prominent approach dates to the sixties, when \cite{KFSeminal} presented the optimal solution when linear dynamical systems with additive white Gaussian noise (AWGN) are considered. Such a filter alternates a prediction phase, also known as time update, where a mathematical model is employed to guess the evolution of the system state, to a measurement update, where the information coming from the observations is incorporated to improve the ``a posteriori'' state knowledge. Since then, many advances have been done in the filtering problem, mostly related to cases where linearity and noise Gaussianity are not granted anymore. For instance, when nonlinear non-Gaussian models are considered, the posterior state distribution can become analytically intractable, hence it is necessary to introduce approximations in order to describe the process uncertainty. In the Assumed Density Filtering (ADF) setting, the predicted or posterior state distributions are approximated by a Gaussian density by means, for instance, of several techniques like model linearization EKF~(\cite{gelb1974book}), sigma-point methods UKF~(\cite{julier2004unscented}), CKF~(\cite{arasaratnam2009cubature}), or particle filtering approaches (\cite{Gordon1993NovelAT}); a detailed survey on distribution moments approximation in the filtering context can be found in \cite{Roth2016NonlinearKF}.

In the past years, the usage of sensors like radars, cameras or lidars has raised the issue of dealing with the phenomenon of clutter, that is the presence of unwanted disturbances when a sensor scan is performed. In terms of tracking, this amounts to receive several observations for a single time step; moreover, it may happen that, even if the target is present in the sensor Field of View (FOV), it goes undetected, hence none of the observations may belong to the object of interest. In order to address such additional challenges, many approaches have been proposed in the literature (\cite{BlackmanPopoliBook,BarShalomBookEstimationandTracking,BarShalom2011Handbook}) where, to achieve the optimal state estimation of one or many unknown targets, it is necessary to consider all the possible associations between the observed data and previous state hypotheses. Such process of Data Association (DA) may spawn a set of new state hypotheses in every filter update; when this happens, the posterior state distribution is represented by a sum Gaussian densities (Gaussian mixture), which can grow unbounded in the number over time, making the corresponding uncertainty representation computationally intractable. When a combinatorial explosion of this kind takes place, it is necessary to introduce approximations to the so obtained Gaussian mixture/sum. The problem of reducing the number of hypotheses in a mixture model falls under the name of Mixture Reduction Problem (MRP), for which many solutions have been proposed in the literature (see \cite{SalmondReduction,Blackman2004,Williams06,Runnalls,PGMR,Crouse,CGMR}). 
Although the mixture reduction can appear as a marginal block in a target tracking scheme,    it represents in fact a fundamental step in the filter recursion, since approximating accurately a given state distribution can impact significantly on the filter robustness and accuracy. 
Nonetheless, the MRP proves to be a very difficult task for many reasons, among which are the lack of closed forms for some computations involved, the computational burden introduced by such routines and what a suitable number of components should be in order to balance model complexity and accuracy (Occam's razor). 
In this work, a recently developed mixture reduction algorithm (\cite{adaCTDReduction}) is considered in order to tackle the mentioned problems and to assess its performances when applied to the tracking of a single object in the presence of clutter. The main goal of this work is first to prove how important the hypotheses reduction step can be for the filter accuracy and robustness, and then to show how an adaptive model selection can represent a good compromise when solving the MRP in the case of target tracking in the presence of clutter.

The work is organized as follows: in Section \ref{sec:ProbForm}, the problem of target tracking in clutter is formulated together with mixture reduction. In Section \ref{sec:PropSol}, a mixture reduction pipeline is proposed in order to solve efficiently and intuitively the MRP. In Section \ref{sec:NumTests}, several numerical tests are reported to both assess the performances of the proposed solution and to raise several points about the importance of a sound mixture reduction routine. Conclusions follow.

\subsection{Notation}
In this work, all the quantities will be defined when required, but some preliminary notation is required in any case. $\Real^n_+$ denotes the set of non-negative vectors in $\Real^n$, $I_n$  is the identity matrix in $\Real^{n\times n}$ while $\ones_n$ denotes a vector of ones in $\Real^n$.
$S_{++}^d\subset\Real^{d\times d}$ denotes the open cone of symmetric positive definite (SPD) $d\times d$ matrices.

\section{Problem Formulation}\label{sec:ProbForm}
\subsection{Target Tracking in Clutter}\label{subsec:TTIC}
The problem of tracking a single target in the presence of missed detections and false alarms (FAs) (called clutter in the tracking literature) inherently leads to mixtures appearing in the posterior densities. We here consider the case when the target state, denoted $x_k$, is modeled to evolve according to the following linear Gauss-Markov model.  
\begin{align}
    x_k=Fx_{k-1}+w_{k-1},
\end{align}
where $x_0\sim\nu(x_0\vert\boldsymbol{0},\Sigma_0)$ and $w_k\sim\nu(w_k\vert\boldsymbol{0},Q)$ is the white process noise with the notation $\nu(\cdot\vert \mu, \Sigma)$ denoting the Gaussian density with the mean $\mu$ and the covariance $\Sigma$. The target originates a measurement $z_k$ in the sensor report with probability $P_D$, which is modeled as
\begin{align}
    z_k=Hx_{k}+v_{k},
\end{align}
where $v_k\sim\nu(v_k\vert\boldsymbol{0},R)$ is the white measurement noise. The number of FAs is distributed according to a Poisson distribution with the mean value $\lambda_c$ and the spatial distribution of the FAs is uniform over the surveillance region.

In a Bayesian framework the aim of the tracker at each time step $k$ is to estimate the posterior distribution $p(x_k|Z_{0:k})$ of the state $x_k$ given all of measurements $Z_{0:k}=\{Z_0,Z_1,\cdots,Z_k\}$, where $Z_i=\{z_i^j\}_{j=1}^{m_i}$ denotes the set of measurements collected from the sensor at time $i$ and $m_i\ge 0$ is the number of measurements collected from the sensor at time $i$. At the time step $k$ the following $m_k+1$ association hypotheses can be formed about origin of the measurements in $Z_k$.
\begin{align}
    \mathcal{H}_k^0:&\,\, \text{All measurements in $Z_k$ are FA.}\\
    \mathcal{H}_k^j:&\,\, \text{The $j$th measurement $z_k^j$ belongs to the target and}\nonumber\\
    &\text{the other measurements in $Z_k$ are FA.}
\end{align}
for $j=1,\ldots,m_k$. Given an hypothesis sequence/history $\mathcal{H}_{0:k}=\{\mathcal{H}_0,\ldots,\mathcal{H}_k\}$, hypothesis conditioned state posterior distribution $p(x_k|Z_{0:k},\mathcal{H}_{0:k})$ can be calculated using a Kalman filter as a Gaussian distribution using the measurements associated to the target in the hypothesis sequence $\mathcal{H}_{0:k}$. The overall posterior state distribution can be calculated as a mixture as follows.
\begin{align}\label{eq:posteriorMHT}
    p(x_k|Z_{0:k})=\sum_{\mathcal{H}_{0:k}} p(x_k|Z_{0:k},\mathcal{H}_{0:k})P(\mathcal{H}_{0:k}|Z_{0:k})
\end{align}
where $P(\mathcal{H}_{0:k}|Z_{0:k})$ denotes the posterior probability of the hypothesis sequence $\mathcal{H}_{0:k}$. Unfortunately the number of possible hypothesis sequences, which is $\prod_{i=0}^k (m_i+1)$, and hence the number of components in the mixture above, increases exponentially and hence one must resort to mixture reduction algorithms to keep the storage and computational requirements at a reasonable level.

The classical target tracking methods differ in how they handle the exponentially growing number of association hypotheses. The single hypothesis tracking approaches reduce the number of association hypotheses to unity at each time step by either pruning all hypotheses except the ``best'' one like the nearest and the strongest neighbor filters~(\cite{LiBarShalom1996,Li1998}), or by merging all association hypotheses into a single composite hypothesis like the probabilistic data association filter (PDAF)~(\cite{BarShalomTse1975,BarShalomDH2009}). Single hypothesis tracking approaches are effective in high SNR environments under low to moderate clutter. For medium SNR or dense clutter environments, Multiple Hypotheses Trackers (MHT)~(\cite{SingerSH1974,Reid1979,Blackman2004}) which can keep and propagate multiple association hypotheses are preferred. In an MHT, an effective mixture reduction scheme is essential to keep the number of association hypotheses under control.

\subsection{Mixtures of Gaussians}
A Gaussian Mixture (GM) is a mixture model defined as:
\begin{align} \label{eq:MoG}
p(x\vert \Theta) = \wbold^T \nubold(x\vert \thetabold) = \sum_{i=1}^n w_i \nu(x\vert \mu_i,\Sigma_i) = \sum_{i=1}^n w_i \nu_i,
\end{align}
where $n$ is the size of the mixture, $\nubold$ is a vector of Gaussian pdfs,
i.e.\  $\nubold=[\nu_1,...,\nu_n]^T$, $\nu_i=\nu(\cdot\vert \mu_i, \Sigma_i)$ is the $d$-dimensional Gaussian density of parameters $\mu_i\in \Real^d$ (mean value) and $\Sigma_i\in S_{++}^d$ (covariance matrix). $\thetabold=(\mubold,\Sigbold)\in(\Real^d\times S_{++}^d)^n$ is the collection of all the GM means and covariances, with $\theta_i=(\mu_i,\Sigma_i)\in(\Real^d\times S_{++}^d)$. $\wbold = [w_1,...,w_n]^T$ is a vector of \textit{weights} belonging to the standard $(n-1)$-dimensional simplex
$\Delta^{n-1}=\{\wbold\in [0,1]^n:\  \wbold^T \ones_n=1\} \subset \Real_+^n$,
so that the mixture $p(x\vert\Theta)$ defined in \eqref{eq:MoG} is
a \textit{convex sum} of the $n$ components $\nu_i$. Finally, $\Theta =(\wbold,\thetabold) = (\wbold, \mubold, \Sigbold)\in
\Bcal_{n}$ is the collection of all mixture parameters, with $\Bcal_{n}=\Delta^{n-1}\times (\Real^d\times S_{++}^d)^n$.

\subsection{The Mixture Reduction Problem}\label{subsec:MRP}
\subsubsection{Kullback-Leibler Divergence.} For the sake of discussion, it is necessary to define a way to quantify how dissimilar two distributions are. In the literature, there exist several dissimilarity measures ($D$-measures, for short) between distributions (\cite{FPI}), but for the goal of this work we will restrict to the \textit{Forward Kullback-Leibler Divergence} (FKLD) (\cite{KLD}, $\DKL$ for short), also known as \textit{differential relative entropy}; the term \textit{forward} comes from the fact that the KL is not a symmetric measure, and the order in which two distributions are considered can impact significantly on the outcome. 
Given two $d$-dimensional Gaussian distributions $\nu_i(x)$ and $\nu_j(x)$, the $\DKL$ takes the following form:
\begin{equation}\label{eq:KLD}
\begin{aligned}
    &\DKL(\nu_i\Vert \nu_j) = \int \nu_i(x) \log \frac{\nu_i(x)}{\nu_j(x)}\d x\\
    &=\frac{1}{2}(\trace(\Sigma_j^{-1} \Sigma_i) + (\mu_i-\mu_j)^T\Sigma_j^{-1}(\mu_i-\mu_j) - d + \log\frac{\vert \Sigma_j\vert}{\vert \Sigma_i\vert})
\end{aligned}
\end{equation}
and is such that:
\begin{align}
& \DKL(\nu_i\Vert \nu_j)\ge 0,\label{eq:Dnonneg}\\  
& \DKL(\nu_i\Vert \nu_j) = 0\ \Longleftrightarrow\ \nu_i=\nu_j.\label{eq:Dident}
\end{align}
The $\DKL$ is a very sound $D$-measure used in many fields and applications, since it possesses a long list of useful properties (it is strictly linked to concepts like \textit{Maximum Likelihood Estimation} (MLE) and information gain; see \cite{pml1Book}).

\subsubsection{Approximating a Mixture Model.}
Given a mixture $p^a=(\wbold^a)^T\qbold^a$, of size $n^a$, and a dissimilarity measure $D(\cdot\Vert \cdot)$, one wants to find a reduced mixture $p^b$, of size $n^b<n^a$, which is as close as possible to $p^a$. Formally, one wants to solve the problem (\cite{CGMR}):
\begin{align} \label{eq:MRformulaz}
    {\Theta^b}^* = \argmin_{\Theta^b\in\Bcal_{n^b}} D\big(p(x\vert \Theta^a)\Vert p(x\vert \Theta^b)\big),
\end{align}
This is in general a complex, non-convex constrained nonlinear optimization problem, which does not admit a closed form solution. Moreover, when dealing with mixtures, only very few $D$-measures admit a closed form; for instance, in the $\DKL$ case, it is not possible to evaluate the divergence between mixtures. For this and other reasons, the MRP is usually approached by means of several heuristics, often driven by ease of computation.

A general approach to reduce a mixture model is to perform a \textit{greedy reduction} of the components, where subsets of hypotheses are \textit{pruned} or \textit{merged together} according to some criteria until a desired $n^b$ is reached; such a reduced order model can possibly serve as initialization for a refinement phase, where the reduced mixture parameters are optimized by exploiting the information contained in the original model.
\section{Proposed Approach}\label{sec:PropSol}
As discussed, reducing the complexity of a mixture model can be a very difficult task, given that, often, closed forms to compute key quantities are not available. Nonetheless, in \cite{CGMR}, a MR reduction framework has been presented that only requires two key ingredients: the closed form computation of the dissimilarity of pairs of mixture components, and the easy evaluation of the $D$-barycenter of a set of weighted Gaussian densities. The $D$-barycenter $\nuhat(x\vert \muhat, \Sigmahat)$ of a set of weighted Gaussian densities $\{w_i\nu_i\}_{i=1}^n$ is obtained by solving the following problem (see \cite{FPI}):
\begin{align}
    (\muhat,\Sigmahat) = \argmin_{(\mu,\Sigma)\in \Real^d\times S_{++}^d}\sum_{i=1}^n w_i D(\nu_i\Vert \nu(\cdot\vert \mu, \Sigma)).
\end{align}
\vspace{-0.3cm}
\subsection{Greedy Reduction of Mixture Models}
If the $\DKL$ is considered, the pairwise dissimilarities between Gaussian hypotheses can be computed as in \eqref{eq:KLD}; regarding the $\DKL$-barycenter, it is also known in the literature as \textit{moment matching} or \textit{moment-preserving merge}\footnote{This name comes from the fact that such a way of merging mixture components preserves the first two mixture moments.}; given a set of weighted Gaussian densities $(\wbold,\nubold)=\{w_i \nu_i\}_{i=1}^n$, the corresponding moment matching approximation (or $\DKL$-barycenter) denoted by $\nuhat = \nu(\cdot\vert \muhat, \Sigmahat)$, can be computed as:
\begin{equation}\label{eq:KLDbaryGauss}
\begin{aligned}
& \muhat = \frac{1}{\wbold^T \ones_n}\sum_{i=1}^n w_i \mu_i,\\
& \Sigmahat = \frac{1}{\wbold^T \ones_n}\sum_{i=1}^n w_i(\Sigma_i + (\mu_i-\muhat)(\mu_i-\muhat)^T).
\end{aligned}
\end{equation}
Given two Gaussian hypotheses $\nu_i$ and $\nu_j$, let us define the cost of merging those two components together as follows (see \cite{CGMR}):
\smallskip
\begin{equation}\label{eq:BBound}
B(w_i\nu_i,w_j\nu_j) = w_i\DKL(\nu_i\Vert \nuhat_{i,j}) + w_j\DKL(\nu_j\Vert \nuhat_{i,j})
\end{equation}

where $\nuhat_{i,j}$ is the $\DKL$-barycenter computed as in \eqref{eq:KLDbaryGauss} of the components $i$ and $j$ of a mixture. Note that such costs are symmetric, even if the $\DKL$ is not; this represents a good advantage in terms of computational operations, since, given a mixture model of size $n$, one has to evaluate only $\frac{n(n-1)}{2}$ merging costs. Given these ingredients, it is possible to formulate a greedy reduction algorithm which, at each step, selects for merging the two components $i$ and $j$ associated to the least value of \eqref{eq:BBound}, and replaces $\nu_i$ and $\nu_j$ with their barycenter $\nuhat_{i,j}$ in the current mixture model; as discussed in \cite{CGMR}, such a choice corresponds to minimize an upper bound on the true, yet intractable, $\DKL$ between the mixture before and the one after the merging: such an upper bound is the \textit{Composite Transportation Dissimilarity}, $C_D$ for short. The procedure goes on until the desired number of components $n^b$ is reached. This algorithm is also known in the literature as the Runnalls' algorithm (\cite{Runnalls}), a special case of the framework presented in \cite{CGMR} when the $\DKL$ is considered.

\subsection{Finding the Appropriate Number of Mixture Components}
Although the discussed algorithm results to be very performing, the issue of deciding what is a suitable number of components for the approximated model remains open; such an operation is a \textit{model selection} problem.
In the perspective of the previously mentioned framework, the costs $B(\cdot,\cdot)$ assume an interesting meaning since, as reported in \cite{adaCTDReduction}, they possess useful properties which allow to embed a model selection routine directly into the greedy reduction phase, thus obtaining an adaptive greedy reduction algorithm. In order to do so, it is necessary to define few more quantities.

Given a GM $p^a = \sum_{i=1}^{n^a} w_i^a \nu_i^a$, let us denote the corresponding $\DKL$-barycenter as $\nuhat^a$, let us denote with $m$ the \textit{current} mixture model order in a greedy reduction algorithm, and with $p^{(m)}$ the corresponding mixture; $m$ will be a decreasing integer starting from $n^a$ and ending to $n^b$. In \cite{adaCTDReduction} is reported that it is possible to evaluate the cost of merging all the mixture components into a single hypothesis $\nuhat^a$ as:
\begin{equation}   \label{eq:mDofp}
    c(\nuhat^a\vert p^a) = \frac{1}{(\wbold^a)^T  \ones_n}\sum_{i=1}^{n^a} w_i^a \DKL(\nu_i^a\Vert \nuhat^a).
\end{equation}
In addition, the following quantity can be defined:
\smallskip
\begin{equation}\label{eq:CumRTLsBinequality}
    \Lcaltilde^{(m)} = \frac{\sum_{l=n^a}^{m}\Bbreve^{(l)}}{c({\nuhat^a}\vert p^a)} \in [0,1]
\end{equation}
that is the sum of all the bounds $\Bbreve^{(l)}$, which are the lowest costs \eqref{eq:BBound} computed for the reduced mixture $p^{(l)}$, so that $\Lcaltilde^{(m)}$ is the sum of all the costs associated to the optimal merging actions from $p^a$ to $p^{(m)}$; $\Lcaltilde^{(m)}$ can be proven to be an upper bound on the true, yet intractable, $\DKL$ (scaled by a factor $c(\nuhat^a\vert p^a)$) between the original mixture $p^a$ and its greedily reduced approximation of order $m\leq n^a$. From another perspective, \eqref{eq:CumRTLsBinequality} represents the relative accuracy loss w.r.t. the original mixture model. By providing a threshold $\alpha_{\Lcaltilde}\in [0,1]$\footnote{For the extreme case of $\alpha_{\Lcaltilde}=1$ the reported algorithm is equivalent to perform single hypothesis filtering as the PDAF.} it is possible to halt the greedy reduction during the descent when the prescribed loss threshold is exceeded. 
The adaptive greedy reduction algorithm so obtained is reported in Algorithm \ref{alg:adaCTDgreedy}\footnote{The Runnalls algorithm has the same structure, but the computation of $\Lcaltilde$ and $c(\nuhat^a\vert p^a)$ is not required, since the only halting condition is to reach the desired number $n^b$.}.

\IncMargin{1.5em}
\begin{algorithm2e}[hbtp]
\label{alg:adaCTDgreedy}
\SetAlgoLined
 \KwData{Original GM $p^a$, of size $n^a$, \\ relative loss threshold $\alpha_{\Lcaltilde}$.}
 \KwResult{Reduced GM $p^b$ of size $n^b\leq n^a$.}
  $m:=n^a$, $p^{(m)}:=p^a$, $\Lcaltilde^{(m)}:=0$\;
  Compute $c({\nuhat^a}\vert p^a)$\;
 \While{$\Lcaltilde^{(m)} \leq\alpha_{\Lcaltilde}$}
 { find $(i,j)\in[1\!:m]$:\ $B(w_i^{(m)}\nu_i^{(m)},w_j^{(m)}\nu_j^{(m)})\le 
    B(w_r^{(m)}\nu_r^{(m)},w_s^{(m)}\nu_s^{(m)})$,  $\forall r>s\in[1:m]$ \label{op:adafindij}\;
    $\Bbreve^{(m)} := B(w_i^{(m)}\nu_i^{(m)},w_j^{(m)}\nu_j^{(m)})$\;
 $\Lcaltilde^{(m-1)}:=\Lcaltilde^{(m)} + \frac{\Bbreve^{(m)}}{c({\nuhat^a}\vert p^a)}$\;\smallskip
 \If{$\Lcaltilde^{(m-1)}\leq\alpha_{\Lcaltilde}$}
 {\smallskip $p^{(m-1)}:=p^{(m)}-w_i^{(m)} \nu_i^{(m)} -w_j^{(m)} \nu_j^{(m)}+(w_i^{(m)}+w_j^{(m)})\nuhat_{i,j}^{(m)}$\;}
 $m:=m-1$\;
 }
 $p^b:=p^{(m)}$\;
\caption{Adaptive $C_{\DKL}$-based Greedy reduction Algorithm}
\end{algorithm2e}
\DecMargin{1.5em}

Note that the algorithm here reported only performs merging of pairwise components and it does not take into account actions like pruning. This is due to the fact that, in the framework presented in \cite{CGMR}, merging is the only optimal action to consider when reducing greedily a mixture model. Nonetheless, compared to pruning, merging can be computationally burdensome.

\subsection{Capping Hypotheses}
In several processing systems, computational resources may be limited, hence the number of hypotheses to be maintained in an MHT has to be capped to a maximum number; in this regard, \textit{capping} is an operation often considered when a mixture has to be approximated. Such a procedure can simply be performed by sorting the hypotheses w.r.t. the corresponding weights, and to preserve only the $n^b$ most significant ones. In Algorithm \ref{alg:adaCTDgreedy}, there is no actual guarantee that the resulting mixture will possess a number of hypotheses below the maximum allowed number $n^b$; to address such an issue, one could either perform capping as described, or modify line $6$ of the reported scheme by adding the condition $m\leq n^b$, that is the accuracy threshold has been violated and the maximum allowed number of components has been reached. In other words, even if one is losing more accuracy than the desired one, the merging in the $\DKL$ sense continues until $n^b$ is reached. When this happens, the user should reconsider the filter parameters or allocated resources, since it may happen that, for the process of interest, the chosen $n^b$ may be insufficient. The adaptive reduction here proposed can be exploited to figure out a suitable number of components for the problem of interest.

\subsection{Pruning Hypotheses}
Before starting a merging procedure of mixture components in a target tracking filter, one can first consider pruning low weight hypotheses in order to save computational power. Although this is a very efficient way to reduce the number of components, like capping, it is in fact a \textit{destructive} practice, in the sense that hypotheses are just discarded and the corresponding information is lost, at the opposite of what happens when merging hypotheses in the $\DKL$ sense.
A common way to perform pruning, which we denote here as \textit{Standard Pruning} (SP), is to consider a threshold $\gamma$ on the component weights and to discard all the hypotheses for which $w_i< \gamma$, $i=1,...,n$. 
Alternatively, we propose here a slightly more refined pruning method, namely \textit{Normalized-Weight Pruning} (NWP), where the thresholding is operated on the weights $w_i$ scaled by the square root of the determinant of the corresponding covariance matrix, i.e., the $i$-th component is pruned if
\begin{equation}\label{eq:NWP}
    \tilde{w}_i < \tilde{\gamma}, \qquad \text{where} \quad 
    \tilde{w}_i =\frac{w_i}{\sqrt{\vert \Sigma_i\vert}},
\end{equation}
and $\tilde{\gamma}$ is the chosen threshold for the NWP.
Such a modification is suggested from the fact that the $\DKL$ is an \textit{inclusive}\footnote{With the term inclusive we refer to $D$-measures which, when employed in the barycenter computation for a set of weighted components, provide a covariance that is larger than those of the merged components (covariance spread), rather than neglecting low density regions. For other details on inclusivity and exclusivity of $D$-measures see \cite{minka2005divergence, Consistency}.} $D$-measure, so that the covariance significantly spreads if two remarkably dissimilar hypotheses are merged together. 
Indeed, the merging criterion \eqref{eq:BBound} may happen to select for merging two distant hypotheses with very low weights, generating unlikely hypotheses with rather large covariance; by employing a suitable threshold in the NWP, these unlikely hypotheses are removed.

\section{Numerical Tests and Discussion}\label{sec:NumTests}

Several numerical tests are here reported to assess the performances of the proposed approach and to lay the basis for the related discussion. 
In order to study the effects of pruning of the components of the GM that models the posterior state distribution in the tracking filter, we consider a capping-only reduction scheme that preserves 
the $n^b$ most significant GM components, with $n^b=30$.
This scheme is denoted {\it Capping-30}. In general, for the sake of notation, the target number of components in a greedy reduction algorithm is appended to the end of its name (e.g., Runnalls-5 denotes the Runnalls procedure where the target number of components is $n^b=5$).
In addition to capping, we consider the following reduction pipelines:
\begin{enumerate}
    \item SP $\rightarrow$ Runnalls-5 $ \rightarrow$ NWP%
    \footnote{In the considered scenario, suitable values for the NWP threshold $\tilde{\gamma}$ are in the interval $[10^{-12},10^{-6}]$ (physical unit of $\tilde{\gamma}$ is $\frac{s^2}{m^4}$.)}\\
    \item SP $\rightarrow$ Runnalls-30 $\rightarrow$ NWP\\
    \item SP $\rightarrow$ Adaptive-30 (Algo \ref{alg:adaCTDgreedy}) $ \rightarrow$ NWP.
\end{enumerate}

The experimental scenario considers constant velocity motion and measurement models as in \cite{SalmondReduction}:
\begin{equation} \label{eq:constvelmod}
\begin{cases}
    x_k = F x_{k-1} + w_{k-1}, \\
    z_k = H x_k + v_k,
\end{cases}
\end{equation}
with $x_k = [p_x,p_y,v_x,v_y]$, $w_k\sim \nu(x\vert \boldsymbol{0}, Q)$, and:
\begin{equation}
    F =\! \begin{bmatrix}
        1 & 0 & \Delta t & 0\\
        0 & 1 & 0 & \Delta t\\
        0 & 0 & 1 & 0\\
        0 & 0 & 0 & 1
    \end{bmatrix}, \ G =\! \begin{bmatrix}
        \Delta t^2/2 & 0\\
        0 & \Delta t^2/2\\
        \Delta t & 0\\
        0 & \Delta t
    \end{bmatrix}, \ Q=\sigma_q^2GG^T,
\end{equation}
and $\Delta t = 1 s$. $z_k = [p_x, p_y]$, $v_k \sim \nu(x\vert \boldsymbol{0}, R)$, and
\begin{equation} \label{eq:outmod}
H = \begin{bmatrix}
        1 & 0 & 0 & 0\\
        0 & 1 & 0 & 0
    \end{bmatrix}, \quad R = \sigma_r^2 I_2.
\end{equation}
$\sigma_q$ and $\sigma_r$ have respectively units of $m/s^2$ and $m$. Regarding the MHT, a probability of detection of $P_D=0.9$ has been considered, and the gating probability\footnote{Gating is a practice which consists in considering only the measurements falling in a confidence region of a hypothesis.} has been set to $P_G = 0.999$. For the sake of investigation, a ground-truth trajectory of $K=100$ steps has been generated as follows:
\begin{equation}\label{eq:trajGen}
\begin{aligned}
    & x_0 = [0,0,10,-10],\\
    & x_k=F x_{k-1} + Gu^1,\ u^1=[\ 0.2 \ \ 0.6\,]^T,& \text{for $k\in[1,50]$},\\
    & x_k=F x_{k-1} + Gu^2,\ u^2=[\ \ 0 \ \ \ \ \-2\,]^T,& \text{for $k\in[51,75]$},\\ 
    & x_k=F x_{k-1} + Gu^3,\ u^3=[-3 \ \ \ \ 1\,]^T,& \text{for $k\in[76,100]$}.
\end{aligned}
\end{equation}
The measurements $z_k$ have been generated out of such ground-truth using the model \eqref{eq:outmod} with $\sigma_r^2 = 60$. 
In the filter, the values $\tilde{\sigma}_q^2= 9$ and $\tilde{\sigma}_r^2 = 70$ have been used for the computations of gains to simulate a moderate mismatch of the filter w.r.t.\ the underlying system model. 

\begin{figure*}[hbtp]
    \centering
    \includegraphics[width=\textwidth]{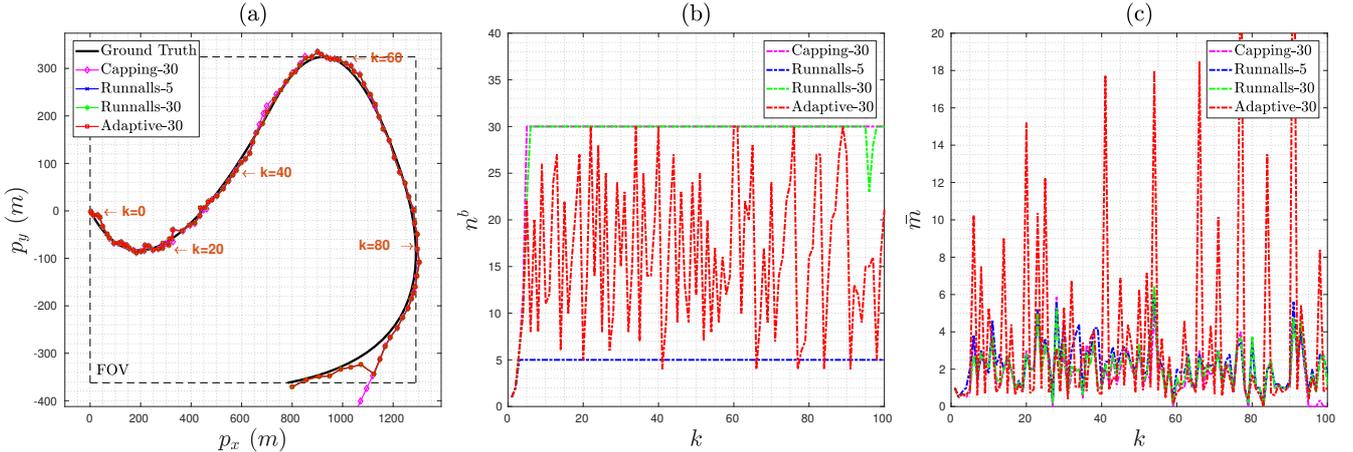}\vspace{-1.4em}
    \caption{(a). Position ground-truth (black, generated as in \eqref{eq:trajGen}) vs. position estimates. (b) Number of maintained hypotheses over epochs. (c) Average measurement number per gate over epochs.}\vspace{-0.5em}
    \label{fig:example}
\end{figure*}

A measurement clutter uniformly distributed over a rectangular FOV of the sensor has been assumed in the performed tests.
The clutter rate (average number of false alarms per scan) in the FOV is denoted $\lambda_c$.
For the sake of comparison, all the results of the numerical tests here reported refer to the same ground-truth \eqref{eq:trajGen} and are carried out by changing the parameters of the various GM reduction schemes
(target mixture size $n^b$, SP threshold $\gamma$, NWP threshold $\tilde{\gamma}$, normalized loss threshold $\alpha_{\Lcaltilde}$)\footnote{Note that the thresholds $\gamma$ and $\alpha_{\Lcaltilde}$ are unitless, while the metric unit of $\tilde{\gamma}$, in the 2D tracking application, is $s^2/m^4$.}
, for different values of the clutter rate $\lambda_c$.
For each filter, the state estimate $\hat{x}_{k\vert k}$ at time step $k$ is computed in the \textit{Minimum Mean Squared Error} (MMSE) sense:
denoting with $\{w_{k\vert k},\mu_{k\vert k,i},\Sigma_{k\vert k,i}\}_{i=1}^n,$
the parameters of the GM of size $n$ that approximates the posterior distribution \eqref{eq:posteriorMHT}, the MMSE state estimate and its covariance are the mean and covariance of the GM, computed as in \eqref{eq:KLDbaryGauss}:
\begin{align}
\hat{x}_{k\vert k} & = \sum_{i=1}^{n} w_{k\vert k,i}\, \mu_{k\vert k,i}\\
\Sigmahat_{k\vert k} & = \sum_{i=1}^n w_{k\vert k,i}
\big(\Sigma_{k\vert k,i} + (\mu_{k\vert k,i}-\hat{x}_{k\vert k} )(\ast)^T\big).
\end{align}
In Fig.\ \ref{fig:example} the outcome of a single experiment is reported, where
$\lambda_c=150$ (FOV area $\approx 700\times 1300\ m^2$),
$\gamma=5\cdot 10^{-4}$, $\tilde{\gamma}=10^{-10}\, s^2/m^4$, and $\alpha_{\Lcaltilde}=0.05$.
By looking at Fig. \ref{fig:example} it is possible to spot how, in the final phase, the Runnalls-based reductions succeed in helping the filter to recover the state track, while the capping-only reduction leads to filter divergence (track loss). 
In Fig.\ \ref{fig:example}.(b) the number of maintained hypotheses over epochs are reported; the capping-only reduction saturates to $n^b=30$ instantly, closely followed by the Runnalls-30 alternative. Nonetheless, the latter undergoes some pruning effects near the end. This is due to the fact that, after recovering the state track, many of the existing hypotheses quickly decreased in the importance (or had very broad covariance), hence falling below the pruning thresholds $\gamma$ and $\tilde{\gamma}$. The Runnalls-5, even if possessing a low number of hypotheses, still succeeds in recovering the track. Finally, the adaptive scheme oscillates a lot below the limit $n^b=30$, while rarely reaching its maximum allowed number. 
In the last plot (Fig.~\ref{fig:example}.(c)) the average number of measurements falling inside the hypotheses gates is reported. 
As it can be observed, less hypotheses obtained by merging may correspond to a higher number of measurements per gate. 
This is due to the fact that the moment-matching merging \eqref{eq:KLDbaryGauss}
while preserving the mixture first and second moments, increases the distribution entropy\footnote{Recall that the Gaussian is the maximum entropy pdf when mean and covariance are given, so that, at the same covariance, mixtures have lower entropy than a single Gaussian.}: 
a higher entropy amounts to more possibilities which, in terms of tracking, may lead to a virtual increase in the uncertainty about the system state (spread of the covariance).
If compared to the capping case, where the uncertainty related to a single hypothesis is the outcome of the Kalman filter only, in the case of schemes where moment-matching takes place, subsequent merging actions increase the overall mixture entropy. Such a phenomenon leads to two main facts:
\begin{itemize}
    \item unlike capping, there is interaction between state tracks, in the sense that merging hypotheses together can be seen as fusing two or more state trajectories into a single one, while increasing the resulting uncertainty. 
    This can enhance the filter robustness to bad modelling or high clutter levels.
    \item In terms of \textit{Root  Mean Squared Error} (RMSE), this may lead to a slightly decrease in the performances when a high clutter is present and aggressive moment-matching is performed. This is due to the fact that reducing significantly the number of hypotheses by merging can lead to larger gates, hence to consider a larger number of data associations mostly associated to clutter; consequently, the corresponding state estimate will be perturbed by the presence of unwanted information. As mentioned, though, this can increase the robustness to bad modelling, thus, in parallel, it could even provide more accurate estimates: the balance between number of maintained hypotheses (model complexity) and approximation methods (model accuracy) may hence be impactful in terms of overall performances. 
\end{itemize}
To assess the previous argumentation, Monte Carlo simulations have been performed for several scenarios, and three metrics have been considered: the position RMSE, the track loss (TL) percentage, and the loop time (LT).
The average number of maintained hypotheses ($\bar{n}^b$) is also reported. 
A track is considered to be lost if, at a given time step, the true state is not contained in any of the hypotheses gates. 
The RMSE is computed only over track sequences which were not lost. 
The loop time considers both the filtering and reduction times
\footnote{ All tests have been performed with a laptop with an Intel Core i7-8750H CPU @ 2.20GHz × 12.
Note that the reported loop times should only have a relative meaning, basically useful to establish an ordering of the algorithms in terms of computational costs.}. 
\subsection{Monte Carlo Tests Over Different Clutter Intensities}
The first test considers a low-to-mid average number of FAs $\lambda_c=150$ over the sensor FOV ($\approx  700\times 1300 m^2)$, for the trajectory reported in \eqref{eq:trajGen}.
With such a rate, the average in-gate measurements for each hypothesis results to be approximately $\bar{m}=3$.
\begin{small}
\begin{table}[hbtp]
\begin{centering}
\begin{tabular}{
|p{1.6cm}||p{1.7cm}|p{1.2cm}|p{1cm}|p{1cm}|}
 \hline
 \multicolumn{5}{|c|}{Monte Carlo average results for $N=1000$ runs} \\
 \hline
 \textbf{Algorithm} & \textbf{RMSE (m)} & \textbf{TL (\%)} & \textbf{LT (s)} & $\boldsymbol{\bar{n}^b}$\\
 \hline
 Capping-30 & 13.1429 & 45.8\% & 0.0060 & 29.2176 \\
 \hline
 Runnalls-5 & 11.9744 & 3.2\% & 0.0059 & 4.9596\\
 \hline
 Runnalls-30 & 11.6724 & 2.0\% & 0.0398& 28.8352\\
 \hline
 Adaptive-30 & 11.7408 & 1.8\% & 0.0162& 15.6205\\
 \hline
\end{tabular}
\smallskip
\caption{Comparison of mixture reduction schemes for $\lambda_c=150$, $n^b=30$, $\gamma=5\cdot10^{-4}$, $\tilde{\gamma}=10^{-10}$, $\alpha_{\Lcaltilde}=0.05$.\label{table:lowClutterMC}\vspace{-1.6em}}
\end{centering}
\end{table}
\end{small}

Table \ref{table:lowClutterMC} shows that, for tracks which were not lost, the Capping-only lags slightly behind in terms of RMSE, while the three other cases are comparable. 
Nonetheless, in terms of track loss, the capping-only scheme has definitely the worst performances. 
For the Runnalls-based alternatives, the filters appear to be significantly more robust. In terms of execution times, the scheme taking up more resources is, as expected, the Runnalls-30, since it involves many cost evaluations and merging actions, while the Runnalls-5 results to be equivalent to the Capping-30 scheme; the Adaptive-30 alternative falls in the middle, suggesting that a suitable number of hypotheses for the process of interest could be $n^b=15$, for the given relative loss threshold $\alpha_{\Lcaltilde}$.

In order to reveal the differences in performance of the considered algorithms, another MC test is reported, where the clutter rate has been increased to $\lambda_c=300$ (Table \ref{table:highClutterMC}). 
\begin{small}
\begin{table}[hbtp]
\begin{centering}
\begin{tabular}{
|p{1.6cm}||p{1.7cm}|p{1.2cm}|p{1cm}|p{1cm}|}
 \hline
 \multicolumn{5}{|c|}{Monte Carlo average results for $N=1000$ runs} \\
 \hline
 \textbf{Algorithm} & \textbf{RMSE (m)} & \textbf{TL (\%)} & \textbf{LT (s)} & $\boldsymbol{\bar{n}^b}$\\
 \hline
 Capping-30 & 21.0946 & 86.9\% & 0.0137 & 29.3432\\
 \hline
 Runnalls-5 & 22.5189 & 21.2\% & 0.0266 & 4.9691\\
 \hline
 Runnalls-30 & 18.4154 & 10.0\% & 0.0869& 29.2683\\
 \hline
 Adaptive-30 & 18.9112 & 10.8\% & 0.0434& 16.0126\\
 \hline
\end{tabular}
\smallskip
\caption{Comparison of mixture reduction schemes for $\lambda_c=300$, $n^b=30$, $\gamma=5\cdot10^{-4}$, $\tilde{\gamma}=10^{-10}$, $\alpha_{\Lcaltilde}=0.05$.\label{table:highClutterMC}\vspace{-1.6em}}
\end{centering}
\end{table}
\end{small}

In terms of RMSE, it is possible to see again how, for tracks which were not lost, a similar ordering as before is maintained; of course, more clutter amounts to a higher RMSE. 
In terms of track loss, one can see how Capping-30 loses the track almost in every run, 
proving its total unreliability as a reduction scheme.
As for the Runnalls-based alternatives, it is more clear now how the number of hypotheses can affect the performances when complex scenarios are addressed: the Runnalls-5 scheme detaches considerably from the other two approaches both in terms of RMSE and TL if compared to the previous case (medium clutter): this could be a first symptom of how aggressive merging can incorporate unwanted information. 
In terms of execution times, we find a similar ordering as for the previous case of $\lambda_c=150$. 
Overall, it seems that a higher number of hypotheses may be necessary if the scenario complexity increases. But to what extent employing a higher number of hypotheses can improve the tracker performances? And to what extent more merging (in the $\DKL$ sense) than pruning can improve the filter robustness? Let us consider, at first, the same experiment as before ($\lambda_c=300$), with the same pruning thresholds, but with $n^b=50$.
\begin{small}
\begin{table}[hbtp]
\begin{centering}
\begin{tabular}{
|p{1.6cm}||p{1.7cm}|p{1.2cm}|p{1cm}|p{1cm}|}
 \hline
 \multicolumn{5}{|c|}{Monte Carlo average results for $N=1000$ runs} \\
 \hline
 \textbf{Algorithm} & \textbf{RMSE (m)} & \textbf{TL (\%)} & \textbf{LT (s)} & $\boldsymbol{\bar{n}^b}$\\
 \hline
 Capping-50 & 20.6443 & 85.8\% & 0.0252 & 48.6863\\
 \hline
 Runnalls-5 & 22.5897 & 21.9\% & 0.0275 & 4.9692\\
 \hline
 Runnalls-50 & 18.6655 & 10.8\% & 0.1577& 48.4460\\
 \hline
 Adaptive-50 & 19.1231 & 11.6\% & 0.0600& 24.0452\\
 \hline
\end{tabular}
\smallskip
\caption{Comparison of mixture reduction schemes for $\lambda_c=300$, $n^b=50$, $\gamma=5\cdot10^{-4}$, $\tilde{\gamma}=10^{-10}$, $\alpha_{\Lcaltilde}=0.05$.\label{table:highClutterMoreHypsMC}\vspace{-1.1em}}
\end{centering}
\end{table}
\end{small}
By looking at Table \ref{table:highClutterMoreHypsMC}, it is possible to see how, w.r.t.\ Table \ref{table:highClutterMC}, not much has changed in terms of RMSE and track loss. 
In this regard, the way hypotheses are managed appears to be more important than the number of maintained hypotheses.
To support this statement, the same test has been repeated ($\lambda_c=300$, $n^b=50$), 
but where the pruning thresholds have been lowered respectively to $\gamma=10^{-4}$ and $\tilde{\gamma}=10^{-12}$. Thus, all algorithms, except for the capping-only scheme, 
perform less pruning and more merging in the $\DKL$ sense.
\begin{small}
\begin{table}[hbtp]
\begin{centering}
\begin{tabular}{
|p{1.6cm}||p{1.7cm}|p{1.2cm}|p{1cm}|p{1cm}|}
 \hline
 \multicolumn{5}{|c|}{Monte Carlo average results for $N=1000$ runs} \\
 \hline
 \textbf{Algorithm} & \textbf{RMSE (m)} & \textbf{TL (\%)} & \textbf{LT (s)} & $\boldsymbol{\bar{n}^b}$\\
 \hline
 Capping-50 & 21.1096 & 86.1\% & 0.0269 &  48.6773\\
 \hline
 Runnalls-5 & 25.6687 & 12.3\% & 0.0994 & 4.9689\\
 \hline
 Runnalls-50 & 20.0868 & 4.0\% & 0.2754& 48.6022\\
 \hline
 Adaptive-50 & 20.2236 & 5.2\% & 0.1036& 21.7148\\
 \hline
\end{tabular}
\smallskip
\caption{Comparison of mixture reduction schemes for $\lambda_c=300$, $n^b=50$, $\gamma=10^{-4}$, $\tilde{\gamma}=10^{-12}$, $\alpha_{\Lcaltilde}=0.05$.\label{table:highClutterMoreHypsLowPruningMC}\vspace{-1.8em}}
\end{centering}
\end{table}
\end{small}

As it can be observed in Table \ref{table:highClutterMoreHypsLowPruningMC}, reducing pruning and increasing merging can improve dramatically the filter robustness, but can worsen the RMSE: as discussed, the Runnalls-5 reduction may be too aggressive, hence by presenting larger gates and enclosing significant noise generated by clutter; in addition, $n^b=5$ can represent in general an insufficient number of hypotheses for the case of interest. Nonetheless, it still presents a significant improvement in robustness if compared to the capping-only scheme. Meanwhile, by employing an adaptive reduction scheme, it is possible to achieve very robust and accurate tracking, close to the Runnalls-50 alternative, while employing a loop time comparable to the Runnalls-5 algorithm. 
A general trend which has been observed in all tests carried out is that Runnalls-based schemes are able to recover the state estimate even if large deviations are taken; at the opposite, capping-only schemes, or alternatives where aggressive pruning is performed, do not possess such property.

\subsection{Discussion}
In this section, we want to discuss further points arising from the numerical tests.
First of all, less hypotheses to be managed do not necessarily imply a lower computational load. 
As mentioned, depending on how the mixture is approximated, the component gates may expand, and considerably more measurements may be enclosed, hence requiring more filter updates in the next time step (see Fig. \ref{fig:example}.(c))%
\footnote{This effect can, however, be mitigated by the NWP \eqref{eq:NWP}.}. 
When this happens, many new hypotheses are generated and more computational burden is added to the reduction phase. 
In this regard, we noted that the performance of the adaptive scheme can be worse if the threshold $\alpha_{\Lcaltilde}$ of the relative accuracy loss is badly tuned. 
Indeed, if a too high value is chosen, aggressive over-reductions may take place, especially in scenarios with low SNR. 
A possible workaround may be to either lower the threshold (hence making the adaptive reduction more \textit{conservative}), or to employ a suitably tuned NWP threshold in order to get rid of those very low importance hypotheses that generate large gates.

The results of many tests performed suggest that the number of hypotheses really useful and significant in a target tracking context, depends on many factors, among which the process dynamics. 
If the target moves essentially in straight lines, it may not be necessary to employ many GM components to achieve robust tracking. 
Conversely, if the process dynamics is more erratic, and not much compatible with the adopted motion model, additional hypotheses could be beneficial to obtain a more robust and accurate tracking.
However, looking carefully at the results of the MC tests, it can be seen that by increasing the number of hypotheses, the average RMSE slightly increases. Again, this is linked to the fact that, if a large number of components is maintained, more tracks associated to clutter are propagated over time, before being pruned or merged, hence perturbing the resulting state estimate. 
In this regard, suitably combining pruning with adaptive reduction can be beneficial. 
We also carried out tests, not reported here, where the ground-truth trajectory was more closely aligned with the transition model \eqref{eq:constvelmod} used for the filter design, and tests with lower noise levels together with low rates of FAs. 
In such cases, often the capping-only scheme has exhibited a decent overall behavior (low track loss percentage and low RMSE), because under the favourable circumstances just described, additional significant hypotheses, other than the one associated to the true state, are unlikely to appear. 
In general, if the knowledge of the process statistics is very accurate (unlikely situation in real world applications), the use of MHTs with many hypotheses may not only be excessive, but in some cases can lead to a deterioration in performance.

In a summary, we have seen that, in some circumstances, an MHT that includes an adaptive GM reduction scheme can achieve essentially the same performances of a Runnalls reduction scheme with a significant number of hypotheses, but with a lower computational load.
But how can it be useful to save computation time, for the same level of performance?
Of course, a shorter filter/reduction loop time would allow increase the sensor scan rate, and in some applications this could be beneficial.
However, in those cases in which there is no need to increase the scan rate, a shorter filter/reduction loop time would allow to carry-out other tasks, useful for further improving the filter performances, like re-estimating the process and measurement noise covariances, or performing a GM refinement routine, as discussed in \cite{CGMR}.

\section{Conclusions and Future Works}\label{sec:Conclusions}

In this work, the problem of tracking a single target in the presence of clutter has been addressed with a special focus on the GM approximation of the posterior distribution. 
After formulating the problem, an approach to adaptively reduce the GM sizes is proposed in order to achieve robust and efficient tracking. 
Several tests carried out have shown that, in target tracking scenarios, where one or more objects have to be tracked in the presence of clutter, the Mixture Reduction is a core component of the filter as a whole, and under some circumstances, if rigorously addressed, can effectively improve the tracking performances. 
The case of multiple-object tracking will be investigated in future works.

\begin{ack}
This work was supported by the Centre of EXcellence on Connected, Geo-Localized and Cybersecure Vehicles (EX-EMERGE), funded by the Italian Government under CIPE resolution n. 70/2017 (Aug.\ 7, 2017).
\end{ack}

\bibliography{ifacconf}             

\begin{thebibliography}{28}
\providecommand{\natexlab}[1]{#1}
\providecommand{\url}[1]{\texttt{#1}}
\providecommand{\urlprefix}{URL }
\expandafter\ifx\csname urlstyle\endcsname\relax
  \providecommand{\doi}[1]{doi:\discretionary{}{}{}#1}\else
  \providecommand{\doi}{doi:\discretionary{}{}{}\begingroup
  \urlstyle{rm}\Url}\fi

\bibitem[{Arasaratnam and Haykin(2009)}]{arasaratnam2009cubature}
Arasaratnam, I. and Haykin, S. (2009).
\newblock Cubature {K}alman filters.
\newblock \emph{IEEE Trans. Automat. Contr.}, 54(6), 1254--1269.

\bibitem[{Bar-Shalom et~al.(2011)Bar-Shalom, Willett, and
  Tian}]{BarShalom2011Handbook}
Bar-Shalom, Y., Willett, P., and Tian, X. (2011).
\newblock \emph{Tracking and Data Fusion: {A} Handbook of Algorithms}.
\newblock YBS Publishing.

\bibitem[{Bar-Shalom et~al.(2009)Bar-Shalom, Daum, and Huang}]{BarShalomDH2009}
Bar-Shalom, Y., Daum, F., and Huang, J. (2009).
\newblock The probabilistic data association filter.
\newblock \emph{IEEE Control Systems Magazine}, 29(6), 82--100.

\bibitem[{Bar-Shalom et~al.(2001)Bar-Shalom, Li, and
  Kirubarajan}]{BarShalomBookEstimationandTracking}
Bar-Shalom, Y., Li, X.R., and Kirubarajan, T. (2001).
\newblock \emph{Estimation with applications to tracking and navigation:
  {T}heory algorithms and software}.
\newblock John Wiley \& Sons.

\bibitem[{Bar-Shalom and Tse(1975)}]{BarShalomTse1975}
Bar-Shalom, Y. and Tse, E. (1975).
\newblock Tracking in a cluttered environment with probabilistic data
  association.
\newblock \emph{Automatica}, 11(5), 451--460.

\bibitem[{Blackman and Popoli(1999)}]{BlackmanPopoliBook}
Blackman, S. and Popoli, R. (1999).
\newblock \emph{Design and Analysis of Modern Tracking Systems}.
\newblock Artech H., Norwood, MA.

\bibitem[{Blackman(2004)}]{Blackman2004}
Blackman, S. (2004).
\newblock Multiple hypothesis tracking for multiple target tracking.
\newblock \emph{IEEE Aerosp. Electron. Syst. Mag.}, 19(1), 5--18.

\bibitem[{Crouse et~al.(2011)Crouse, Willett, Pattipati, and Svensson}]{Crouse}
Crouse, D.F., Willett, P., Pattipati, K., and Svensson, L. (2011).
\newblock A look at {G}aussian mixture reduction algorithms.
\newblock In \emph{14th Int.\ Conf.\ on Information Fusion}.

\bibitem[{D'Ortenzio and Manes(2021)}]{Consistency}
D'Ortenzio, A. and Manes, C. (2021).
\newblock Consistency issues in {G}aussian mixture models reduction algorithms.
\newblock \doi{10.48550/ARXIV.2104.12586}.
\newblock \urlprefix\url{https://arxiv.org/abs/2104.12586}.

\bibitem[{D'Ortenzio et~al.(2022)D'Ortenzio, Manes, and Orguner}]{FPI}
D'Ortenzio, A., Manes, C., and Orguner, U. (2022).
\newblock Fixed-point iterations for several dissimilarity measure barycenters
  in the {G}aussian case.
\newblock \doi{10.48550/ARXIV.2205.04806}.
\newblock \urlprefix\url{https://arxiv.org/abs/2205.04806}.

\bibitem[{D’Ortenzio and Manes(2021)}]{CGMR}
D’Ortenzio, A. and Manes, C. (2021).
\newblock Composite transportation dissimilarity in consistent {G}aussian
  mixture reduction.
\newblock In \emph{IEEE 24th Int.\ Conf.\ on Information Fusion}.

\bibitem[{D’Ortenzio et~al.(2022)D’Ortenzio, Manes, and
  Orguner}]{adaCTDReduction}
D’Ortenzio, A., Manes, C., and Orguner, U. (2022).
\newblock A model selection criterion for the mixture reduction problem based
  on the {K}ullback-{L}eibler divergence.
\newblock In \emph{IEEE 25th Int.\ Conf.\ on Information Fusion}.

\bibitem[{Gelb(1974)}]{gelb1974book}
Gelb, A. (1974).
\newblock \emph{Applied Optimal Estimation}.
\newblock MIT Press.

\bibitem[{Gordon et~al.(1993)Gordon, Salmond, and Smith}]{Gordon1993NovelAT}
Gordon, N., Salmond, D., and Smith, A. (1993).
\newblock Novel approach to nonlinear/non-{G}aussian {B}ayesian state
  estimation.
\newblock \emph{{IEEE} Proc. F Radar Signal Process.}, 140(2), 107--113.

\bibitem[{Huber and Hanebeck(2008)}]{PGMR}
Huber, M.F. and Hanebeck, U.D. (2008).
\newblock Progressive {G}aussian mixture reduction.
\newblock In \emph{2008 11th Int.\ Conf.\ on Information Fusion}, 1--8.

\bibitem[{Julier and Uhlmann(2004)}]{julier2004unscented}
Julier, S.J. and Uhlmann, J.K. (2004).
\newblock Unscented filtering and nonlinear estimation.
\newblock \emph{Proc. IEEE}, 92(3), 401--422.

\bibitem[{Kalman(1960)}]{KFSeminal}
Kalman, R.E. (1960).
\newblock A new approach to linear filtering and prediction problems.
\newblock \emph{Transactions of the ASME--Journal of Basic Engineering},
  82(Series D), 35--45.

\bibitem[{Kullback and Leibler(1951)}]{KLD}
Kullback, S. and Leibler, R.A. (1951).
\newblock On information and sufficiency.
\newblock \emph{The Ann.\ of Math.\ Statistics}, 22(1), 79--86.

\bibitem[{Li(1998)}]{Li1998}
Li, X. (1998).
\newblock Tracking in clutter with strongest neighbor measurements. {I}.
  {T}heoretical analysis.
\newblock \emph{IEEE Trans. Automat. Contr.}, 43(11), 1560--1578.

\bibitem[{Minka(2005)}]{minka2005divergence}
Minka, T. (2005).
\newblock Divergence measures and message passing.
\newblock Technical Report MSR-TR-2005-173, Microsoft.

\bibitem[{Murphy(2022)}]{pml1Book}
Murphy, K.P. (2022).
\newblock \emph{Probabilistic Machine Learning: An introduction}.
\newblock MIT Press.

\bibitem[{Reid(1979)}]{Reid1979}
Reid, D. (1979).
\newblock An algorithm for tracking multiple targets.
\newblock \emph{IEEE Trans. Automat. Contr.}, 24(6), 843--854.

\bibitem[{Rong~Li and Bar-Shalom(1996)}]{LiBarShalom1996}
Rong~Li, X. and Bar-Shalom, Y. (1996).
\newblock Tracking in clutter with nearest neighbor filters: {A}nalysis and
  performance.
\newblock \emph{IEEE Trans. Aerosp. Electron. Syst.}, 32(3), 995--1010.

\bibitem[{Roth et~al.(2016)Roth, Hendeby, and Gustafsson}]{Roth2016NonlinearKF}
Roth, M., Hendeby, G., and Gustafsson, F.K. (2016).
\newblock Nonlinear kalman filters explained: A tutorial on moment computations
  and sigma point methods.
\newblock \emph{Journal of Advances in Information Fusion}, 11, 47--70.

\bibitem[{Runnalls(2007)}]{Runnalls}
Runnalls, A. (2007).
\newblock {K}ullback-{L}eibler approach to {G}aussian mixture reduction.
\newblock \emph{IEEE Trans. Aerosp. Electron. Syst.}, 43(3), 989--999.

\bibitem[{Salmond(1990)}]{SalmondReduction}
Salmond, D.J. (1990).
\newblock {Mixture reduction algorithms for target tracking in clutter}.
\newblock In O.E. Drummond (ed.), \emph{Signal and Data Processing of Small
  Targets 1990}, volume 1305. Int.\ Society for Optics and Photonics, SPIE.

\bibitem[{Singer et~al.(1974)Singer, Sea, and Housewright}]{SingerSH1974}
Singer, R., Sea, R., and Housewright, K. (1974).
\newblock Derivation and evaluation of improved tracking filter for use in
  dense multitarget environments.
\newblock \emph{IEEE Trans. Inf. Theory}, 20(4), 423--432.

\bibitem[{Williams and Maybeck(2003)}]{Williams06}
Williams, J. and Maybeck, P. (2003).
\newblock Cost-function-based {G}aussian mixture reduction for target tracking.
\newblock In \emph{Sixth Int.\ Conf.\ of Information Fusion, 2003. Proceedings
  of the}, volume~2, 1047--1054.

\end{thebibliography}
\end{document}